\documentclass[twocolumn, prb,showpacs]{revtex4-1}
\usepackage{amsmath}
\usepackage{mathrsfs} 
\usepackage{amssymb} 
\usepackage{xcolor,graphicx}
\usepackage{subfigure}
\usepackage{dcolumn}
\usepackage{bm}
\usepackage{natbib}
\usepackage[latin1]{inputenc}

\begin{document}

\title{Strain-tunable magnetic anisotropy in monolayer CrCl$_3$, CrBr$_3$, and CrI$_3$}

\author{Lucas Webster}
\email{lwebst6@students.towson.edu}
\affiliation{Department of Physics, Astronomy, and Geosciences, Towson University, 8000 York Road, Towson, MD 21252, USA}

\author{Jia-An Yan}
\email{jiaanyan@gmail.com}
\affiliation{Department of Physics, Astronomy, and Geosciences, Towson University, 8000 York Road, Towson, MD 21252, USA}

\begin{abstract}
Recent observation of intrinsic ferromagnetism in two-dimensional (2D) CrI$_3$ is associated with the large magnetic anisotropy due to strong spin-orbit coupling (SOC) of I. Magnetic anisotropy energy (MAE) defines the stability of magnetization in a specific direction with respect to the crystal lattice and is an important parameter for nanoscale applications. In this work we apply the density functional theory to study the strain dependence of MAE in 2D monolayer chromium trihalides CrX$_3$ (with X = Cl, Br, and I). Detailed calculations of their energetics, atomic structures and electronic structures under the influence of a biaxial strain $\varepsilon$ have been carried out. It is found that all three compounds exhibit ferromagnetic ordering at the ground state (with $\varepsilon$=0) and upon applying a compressive strain, phase transition to antiferromagnetic state occurs. Unlike in CrCl$_3$ and CrBr$_3$, the electronic band gap in CrI$_3$ increases when a tensile strain is applied. The MAE also exhibits a strain dependence in the chromium trihalides: it increases when a compressive strain is applied in CrI$_3$, while an opposite trend is observed in the other two compounds. In particular, the MAE of CrI$_3$ can be increased by 47\% with a compressive strain of $\varepsilon$ = 5\%.

\end{abstract}

\maketitle


\section{Introduction\label{intro}}

One of the latest advances in the field of two-dimensional (2D) materials is the observation of intrinsic ferromagnetism in monolayers of CrGeTe$_3$\cite{Gong2017} and CrI$_3$\cite{Huang2017}. These systems provide an exciting platform for studying the interplay between various competing electronic and magnetic phenomena at nanoscale, when quantum confinement condition is included. These include, for example, magnetoelectric effect\cite{JiangS2018,Jiang2018,Huang22017,ShengweiJiang2018}, spin/valley physics \cite{Zhong2017} and light-matter interactions under the influence of magnetic ordering \cite{Seyler2017}.

Unlike in bulk magnetic materials, the long-range magnetic ordering in 2D structures is impossible without magnetic anisotropy, which is required for counteracting thermal fluctuations\cite{Mermin1966}. Therefore, magnetic anisotropy, which originates mainly from spin-orbit coupling (SOC) effects\cite{Lado2017}, becomes an important parameter when it comes to 2D magnets as it is qualitatively related to their magnetic stability. Moreover, ferromagnetic 2D materials with large magnetic anisotropy are of great interest for high density magnetic memories and spintronic applications at nanoscale, as in spin valves and magnetic tunnel junctions\cite{Gurney2005,Kryder1992,Prinz1998}.

Strain engineering has been shown to be an effective approach to tune properties of nanomaterials by using the substrate lattice mismatching \cite{Conley2013,ZhangP2016,Peng2014}. It has been demonstrated that external strain can tune the electronic energy band gap in single layer MoS$_{2}$ \cite{Conley2013}. In particular, the way in which strain can affect magnetic anisotropy has been subject of several studies involving thin films \cite{Heuver2015,Rajapitamahuni2016} and 2D materials from density-functional theory (DFT) calculations \cite{Zhang2016,Zhuang2016,ZhangJ2017}.

Typically 2D crystals can sustain larger strains than their bulk counterpart \cite{Bertolazzi2011,Kim2009}. Single layer MoS$_{2}$ can sustain strains as large as 11\% \cite{Bertolazzi2011}, and 6\% in single layer FeSe \cite{ZhangP2016,Peng2014}. DFT calculations show that monolayer chromium trihalides are soft when compared with other 2D materials. A 2D Young's modulus of 24, 29 and 34 N/m has been reported for CrI$_{3}$, CrBr$_{3}$ and CrCl$_{3}$ respectively \cite{Zhang2015}. This is much smaller than that of graphene (340 N/m) \cite{Politano2015}, monolayer MoS$_{2}$ (180 N/m)\cite{Bertolazzi2011} and monolayer FeSe (80 N/m) \cite{Zheng2017}.

The softness of single layer chromium trihalides implies that strain modulation of their electronic and magnetic properties can be realized in these systems. Here we provide an extensive study on structural modification at nanoscale resulting from biaxial strain, by means of first-principle calculations. We also investigate the electronic structures, magnetism and magnetic anisotropy of the single layer chromium trihalides under different strains.

The paper is organized as follows. In Section~\ref{method}, the calculation details are given. In Section~\ref{result}, we discuss our results. A brief conclusion will be drawn in Section.~\ref{conclusion}.

\begin{figure}[tbp]
\centering
\includegraphics[width=8.5 cm]{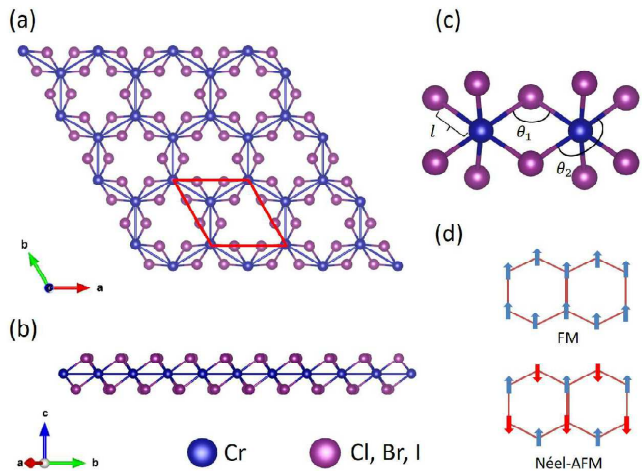}
\caption{\label{fig:model}Atomic structure of monolayer CrI$_3$. (a) Top view and (b) side view of a single layer; (c) Bonding between chromium and iodine atoms. The unit cell of CrI$_3$ which includes two Cr and six I atoms has been indicated in (a). The bond length $l$ between Cr and I atom, the bond angle $\theta_1$ between Cr and two I atoms in the same plane, and the axial angle $\theta_2$ are also shown in (c). The two magnetic orders, namely Néel-antiferromagnetic (AFM) and ferromagnetic (FM) are displayed in (d) }
\end{figure}

\section{Calculational Methods \label{method}}

Our DFT calculations were performed using projected augmented wave (PAW) method as implemented in the Vienna ab initio Simulation Package (VASP)\cite{Kresse1996,Kresse1999}. In all of these calculations we adopted the Perdew-Burke-Ernzerhof (PBE) \cite{Perdew1996} flavor for the generalized-gradient exchange-correlation functional. The Brillouin zone was sampled by 8$\times$8$\times$1 $k$-point grid mesh \cite{Monkhorst1976}, and a 500 eV plane wave cutoff energy was used. Moreover, a 15 \AA\ vacuum was applied along the $z$ axis to avoid any artificial interactions between images. Relaxations were performed until the Hellmann-Feynman force on each atom becomes smaller than 0.002 eV/\AA\ and the total energy was converged to be within $10^{-8}$ eV. Spin-polarization had to be taken into account in order to reproduce the semiconducting nature of this system, and the effect of introducing SOC on the electrical and mechanical properties of these materials will be discussed. In addition, two different magnetic configurations were considered to evaluate the magnetic ground state by comparing their total energies. The ferromagnetic (FM) configuration had all magnetic moments initialized in the same direction while in the antiferromagnetic (AFM) configuration the magnetic moments were set to be antiparallel between nearest neighbors. For both cases, spin orientations were initially in the off-plane direction. These two typical magnetic orderings are shown in Fig.~1(d).

Magnetic anisotropy energy (MAE) is defined as the difference between energies corresponding to the magnetization in the in-plane and off-plane directions (MAE = $E_\parallel-E_\perp$). Therefore, the positive (negative) value of MAE indicates off-plane (in-plane) easy-axis. To evaluate MAE, one must take SOC effects into account. Thus, non-collinear non-self-consistent calculations were performed to evaluate the total energies $E_\parallel$ and $E_\perp$ after the self-consistent ground states were achieved.

In this work, only biaxial strain has been studied. For each strain, the lattice constants were changed accordingly and then kept fixed, while the atomic positions were fully optimized. This procedure was repeated for single layer CrCl$_{3}$, CrBr$_{3}$ and CrI$_{3}$.

\section{Results and Discussions}\label{result}

\subsection{Atomic structures}
\begin{table*}[tbp]
\centering
 \caption{The calculated lattice constants a$_0$, total energy E$_t$, bond lengths $l$, bond angles $\theta_1$ and axial bond angles $\theta_2$ for the FM phase of the chromium trihalides. The difference in energy between two magnetic phases E$_{FM}$-E$_{AFM}$, magnetic anisotropy energy (MAE) and the easy magnetization axis are also listed. SOC has been included in all calculations.} \label{tab2}
\begin{ruledtabular}
\begin{tabular}{lllllllll}
  & \multicolumn{5}{c}{Lattice Parameters}  & \multicolumn{3}{c}{Magnetic Stability for 1L}  \\
    \cline{2-6}    \cline{7-9}
   & a$_0$ (\AA) & E$_t$ (eV) & $\theta_1 (^{o})$ & $\theta_2 (^{o})$ & $l$ (\AA) & E$_{FM}$-E$_{AFM} (eV)$ & MAE ($\mu eV/Cr$) & Easy Axis \\
   \hline
CrCl$_{3}$	& 6.056	&	-38.916  &	95.8	& 173.2  &	2.357	& -0.023	&	24.68	&c\\
CrBr$_{3}$	& 6.438	&   -35.334  &	95.1	& 173.5  &	2.518	& -0.032	&	159.54	&c\\
CrI$_{3}$ 	& 7.008	&	-32.318  &	95.2	& 173.3  &	2.740	& -0.036	&	803.65	&c\\
\end{tabular}
\end{ruledtabular}
\end{table*}

The bulk chromium trihalides CrX$_{3}$ (X=Cl, Br, and I) are layered van der Waals (vdW) materials and the possibility of mechanically exfoliating CrI$_{3}$ to produce 2D monolayers has been demonstrated\cite{Huang2017}. These systems exhibit rhombohedral BiI$_{3}$ structure (space group $R\bar{3}$) in cryogenic temperatures at which ferromagnetism can be observed. The Curie temperatures for the bulk systems are 27, 47, and 70 K for CrCl$_{3}$, CrBr$_{3}$, and CrI$_{3}$ respectively \cite{McGuire2015,McGuire2017}. In the single layer limit, CrI$_{3}$ retains its ferromagnetism, and the Curie temperature of the 2D system is found to be $T_c$ = 45 K\cite{Huang2017}. Fig.~1(a) shows a schematic plot of the single layer chromium trihalide compound. The chromium ions form a honeycomb network sandwiched by two atomic planes of halide atoms as shown in Figs.~1(a) and 1(b). The parallelogram in Fig.~1(a) represents the unit cell containing two chromium atoms and six halide atoms per layer. Moreover, Cr$^{3+}$ ions are coordinated by edge-sharing octahedra, as shown in Fig.~1(c).

Magnetism in these compounds arises from the partially filled $d$ orbitals, as Cr$^{3+}$ ion has an electronic configuration of 3d$^{3}$. Despite this fact, these materials are found to be electrical insulators, indicating that a Mott-Hubbard mechanism is playing a role in the formation of the band gap. In the octahedral environment, crystal field interaction with the halide ligands results in the quenching of orbital moment ($L$ = 0) and splitting of the chromium $d$ orbitals into a set of triply degenerate t$_{2g}$ orbitals (with lower energy) and doubly degenerate e$_{g}$ orbitals (with higher energy). Furthermore, saturation magnetization measurements give an atomic magnetic moment of 3 $\mu_B$ per chromium atom \cite{Richter2018}. This is consistent with Hund's rule which predicts that the three electrons will occupy the t$_{2g}$ triplet yielding $S$ = 3/2.

First, we investigate the structure, magnetism and MAE of the unstrained monolayer systems. Our results from non-collinear self-consistent calculations show that the ground state for the three systems is FM, as indicated in Table 1. Here, the difference in the total energy of the two magnetic phases considered in our study (E$_{FM}$ - E$_{AFM}$) is less than zero for all the chromium trihalides. For this reason, only the optimized structural parameters for the FM ground state are shown in Table 1. For example, for the case of CrI$_{3}$, the lattice constant ($a_{0}$) and Cr-I bond length ($l$) are 7.008 \AA\ and 2.740 \AA\ respectively. These results are consistent with previously reported values \cite{Zhang2015, McGuire2015}. The Cr-I-Cr bond angle ($\theta_1$) is 95.2º and is represented in Fig.~1(c). This angle accounts for the ferromagnetic superexchange interaction according to the Goodenough\cite{Goodenough1958} and Kanamori\cite{Kanamori1958} rules. The axial angle ($\theta_2$) is the angle formed between the chromium ion and two opposing ligands within the same octahedral (e.g. Fig.~1(c)). Therefore, our results suggest some deformation in the octahedral environment as the axial angle is predicted to be slightly smaller than 180$^\circ$. Interestingly, these angles are nearly independent from the ligands, as they are roughly the same for all three systems. In our previous work, we found that different magnetic orderings (FM or AFM) yield similar results for the structural parameters \cite{Webster2018}. Moreover, these structural parameters are not sensitive to the SOC \cite{Webster2018}, which is essential for investigating the MAE below.

The calculated MAE for each compound is listed in Table~1. The MAE of CrI$_{3}$ is about 804 $\mu$eV/Cr, surprisingly large when compared to the other compounds in Table 1. Previous studies reported 980 $\mu$eV/Cr \cite{Jiang2018} and 686 $\mu$eV/Cr \cite{Zhang2015}, and these discrepancies might result from different methods adopted. Lado \emph{et al.} suggested that the large MAE in CrI$_3$ is originated from an anisotropic exchange interaction through a superexchange mechanism, which stems from the strong SOC in the heavier iodine ions \cite{Lado2017}. In addition, the easy axis for energetically favorable spontaneous magnetization is found to be perpendicular to the basal plane, i.e., along the $c$ direction.

\subsection{Electronic structures}

\begin{figure*}[tbp]
\centering
\includegraphics[width=1.00\textwidth]{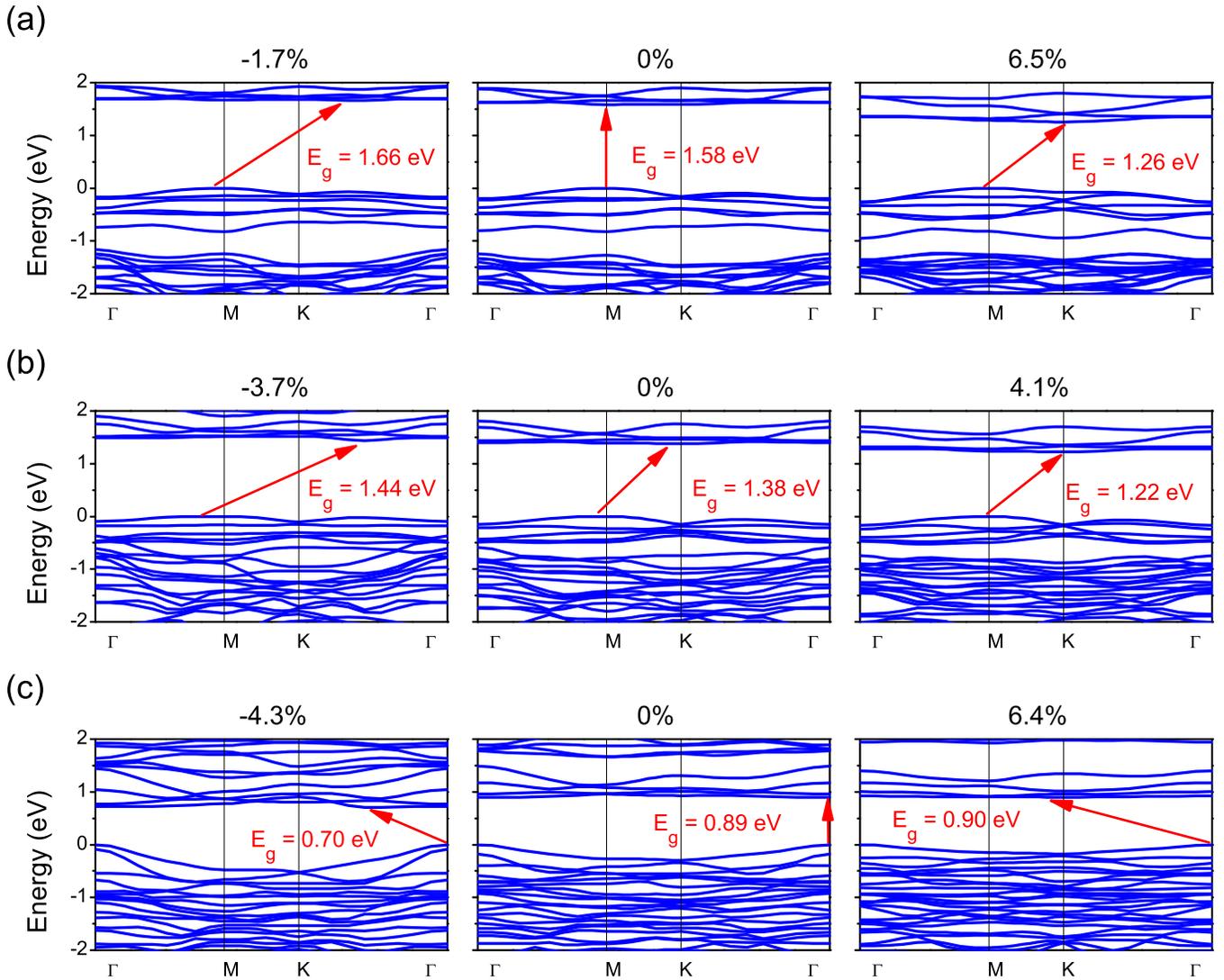}
\caption{\label{fig:band2}Non-collinear spin polarized electronic band dispersions for monolayer chromium trihalides. The effect of typical compressive and tensile strain are also shown in (a) CrCl$_{3}$; (b) CrBr$_{3}$; (c) CrI$_{3}$. The energy band gaps between conduction band minimum (CBM) and valence band maximum (VBM) have been indicated using red arrows in each case. The VBM has been shifted to zero.}
\end{figure*}

The electronic band structures for CrCl$_{3}$, CrBr$_{3}$ and CrI$_{3}$ under various biaxial strain are shown in Figs.~2(a), 2(b) and 2(c), respectively. In Ref.\cite{Webster2018}, we already show that the band structure of CrI$_{3}$ is highly sensitive to the magnetic ordering, the exchange-correlation functional, and SOC. Here we focus on the results calculated from PBE for the FM phase only. The effect of SOC on the electronic band structure can be further elucidated by comparing Fig.~2 with Fig.~S1 in the Supplemental Materials \cite{sup2018}, which shows the collinear results. Unlike CrI$_{3}$, the band structures of the other compounds are insensitive to the inclusion of SOC. This provides a further evidence that MAE in these systems is closely related to the SOC in the halide ligands.

Clearly, the band gaps increase from 0.89 eV to 1.58 eV from iodine to chlorine. Both CrCl$_{3}$ and CrI$_{3}$ exhibit a direct band gap. The band gap character found for unstrained CrI$_{3}$ is consistent with previous works, which also included SOC \cite{Jiang2018, Zheng2017}. In addition, from Fig.~2(a) and 2(b), we notice an increase (decrease) in the band gap upon compression (tensile strain) in these systems. However, according to Fig.~2(c), the application of a biaxial strain causes an opposite effect on the band gap of CrI$_{3}$. In the three compounds, the VBM remains approximately constant with the application of strain, whereas the CBM is shifted, causing a direct-to-indirect band gap transition in both CrCl$_{3}$ and CrI$_{3}$. Moreover, a compressive strain will decrease the energy of the valence bands near $M$ and $K$ points in CrI$_{3}$. This effect can also be observed in collinear calculations, however it is more evident when SOC is included.


\subsection{Effect of Strain on Crystal Structure and the Magnetic Order}

Next, we investigate the dependence of magnetic properties under different strains. It is mainly Cr atoms that contribute to the magnetic moment, which remains overall constant with 6 $\mu_B$ (two Cr$^{3+}$ ions per unit cell) per unit cell under strain (not shown). Fig.~3 shows the energy difference between the two magnetic orderings, namely FM and AFM. A phase transition from FM to AFM is observed in all systems and the AFM region is highlighted in red. Here, the strain $\varepsilon$ is defined as following:

\begin{equation}
\varepsilon =\frac{\left ( a-a_{0} \right )}{a_{0}},
\end{equation}
where $a_{0}$ is the lattice constant for the unstrained system. The energy difference of the two phases is given by (neglecting the MAE since it is relatively small): \cite{Zhang2015}
\begin{equation}
E_{FM/AFM} = E_{0}-\left ( \pm 3J_{1} + 6J_{2} \pm 3J_{3} \right )\left | \vec{S} \right |^{2},
\end{equation}
where $J_{1}$, $J_{2}$ and $J_{3}$ are the Heisenberg exchange integrations for the first, second and third nearest neighbors respectively. Neglecting the second and third nearest neighbors and taking the energy difference between FM and AFM phases, we have:

\begin{equation}
E_{FM}-E_{AFM}=-6J\left | \vec{S} \right |^{2}.
\end{equation}
Here $\left |\vec{S} \right|=3/2$. From the energy difference calculated in DFT (as shown in Table I), one can determine the exchange parameter $J$ with Eq.~3. Next, with $J$ available one can roughly estimate the Curie temperature from the mean-field expression:
\begin{equation}
T_{c}=\frac{3J}{2K_{B}}.
\end{equation}

\begin{table}[tbp]
\centering
 \caption{The exchange coupling and Curie temperature of single-layer chromium trihalides.} \label{tab2}
\begin{ruledtabular}
\begin{tabular}{llll}
   & J (meV) & Estimated T$_c$ (K) & Experimental T$_c$ (K) \\
   \hline
CrCl$_{3}$	& 1.7 &	29.7 & 27 (Bulk)\cite{McGuire2015} \\
CrBr$_{3}$	& 2.4 &	41.3 & 47 (Bulk)\cite{McGuire2015}\\
CrI$_{3}$ 	& 2.7 &	46.4 & 70 (Bulk)\cite{McGuire2015}, 45 (1L)\cite{Huang2017}\\
\end{tabular}
\end{ruledtabular}
\end{table}

A brief derivation is provided in the Supplemental Materials \cite{sup2018}. The Heisenberg exchange parameter and the estimated Curie temperature for the unstrained systems are listed in Table II, along with the available experimental values for the Curie temperature. The obtained value for $J$, considering only the first nearest neighbors, agrees with previous DFT calculations \cite{Zhang2015}. From Table~\ref{tab2}, the estimated Curie temperatures for single layer CrCl$_{3}$ and CrBr$_{3}$ are approximately equal to the experimental values for the bulk systems. Whereas there is a difference between the Curie temperatures for the bulk CrI$_{3}$ and monolayer CrI$_{3}$, our estimation is closer to that of monolayer CrI$_{3}$. This is expected, since Eq.~4 is valid for the 2D system, and suggests a relatively large interlayer coupling in bulk CrI$_{3}$.

In Fig.~3, we show the energy difference $E_{FM}-E_{AFM}$ between the two magnetic ordering as a function of $\varepsilon$ for CrCl$_3$, CrBr$_3$ and CrI$_3$. The evolution of the Curie temperature calculated based on Eq.~(4) has been also shown in Fig.~3 (right axis). As $\varepsilon$ decreases (i.e., the compressive strain increases), the energy difference increases. There is a phase transition to the AFM phase when the energy difference between FM and AFM ordering becomes greater than zero. This phase transition occurs at -2.5 \%, -4.1 \% and -5.7 \% compressive strain for the CrCl$_{3}$, CrBr$_{3}$ and CrI$_{3}$, respectively. Similar result was reported by Zheng \emph{et al.} for CrI$_{3}$ under compressive strain\cite{Zheng2017}. According to Fig.~3, there is a point in which the Curie temperature is maximum and this point is close to the equilibrium for CrI$_{3}$. However, as shown in Fig.~3(a), this point is slightly shifted from the equilibrium to the region with small tensile strain in CrCl$_{3}$ and CrBr$_{3}$, suggesting that in these cases the Curie temperature can be further increased with the introduction of tensile strain. We estimate that a tensile strain of 2.4 \% can increase T$_c$ to 39.0 K in CrCl$_{3}$, and a tensile strain of 2.1 \% can increase T$_c$ to 44.4 K in CrBr$_{3}$. Beyond this point of local maximum, the Curie temperature tends to decrease and no transition to the AFM phase is observed if we further increase tensile strain within the 10 \% range.

We have also calculated the effect of strain on the structural parameters as depicted in Fig.~4. These calculations were repeated for each material with different magnetic ordering (FM and AFM). Since results of AFM are similar, we show only the results for the FM phase. From Fig.~4, one can clearly see that the Cr-X-Cr angle changes linearly with respect to strain. In addition, the angle with $\varepsilon$ = 0 is roughly the same (95$^\circ$) for the three halides. Within the 10 \% range from the equilibrium point ($\varepsilon$ = 0), the curves display a linear shape with the same slope for the three materials. The axial angle has a similar behavior, but tends to increase upon compression until it saturates near 180$^\circ$. At this point, with the axial angle fully stretched (around -10 \% of compression for all materials) the system achieves a higher degree of symmetry. The Cr-X distance displays a weak strain dependency within the 10 \% range.
\begin{figure}[tbp]
\centering

\includegraphics[height=16.0 cm, clip]{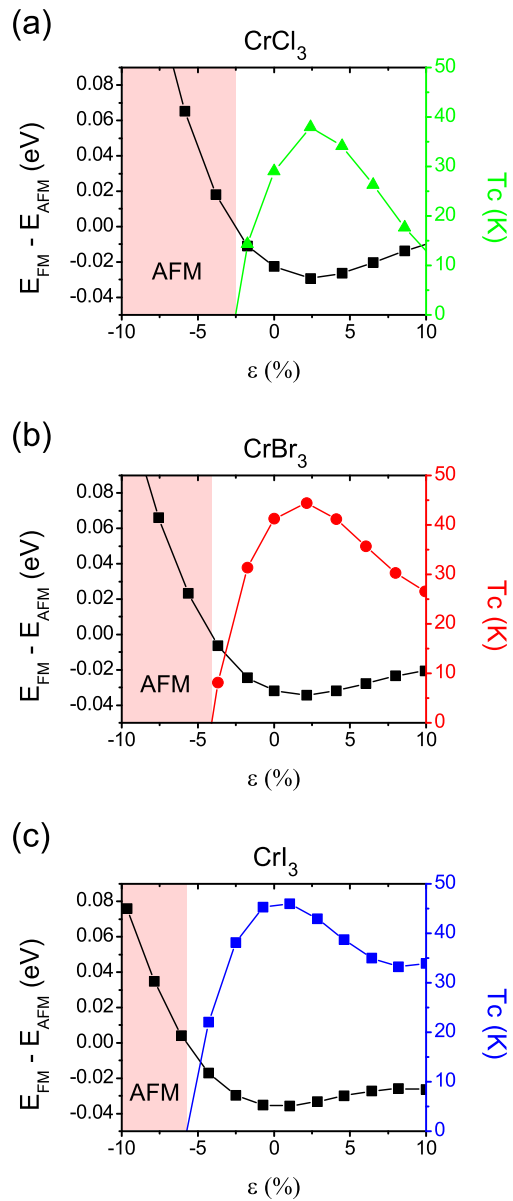}
\caption{\label{fig:magord}Energy difference between FM and AFM phases for (a) CrCl$_{3}$; (b) CrBr$_{3}$; (c) CrI$_{3}$. The AFM phase region is highlighted in red. The calculated Curie temperature is also shown for each case. }
\end{figure}

\begin{figure*}[tbp]
\centering
\includegraphics[height=5.5 cm, clip]{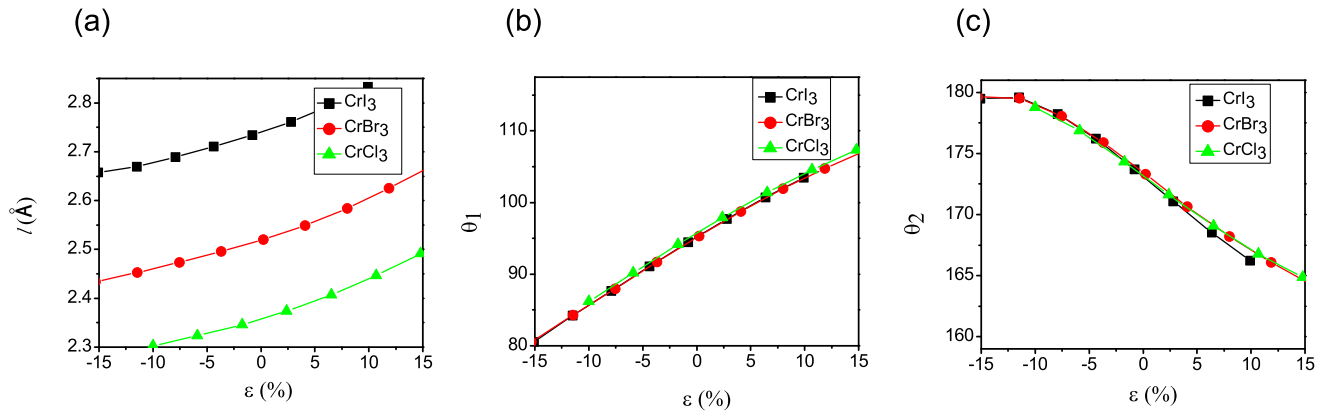}
\caption{\label{fig:str}Dependence of structural parameters on biaxial strain $\varepsilon$ for the chromium trihalides. (a) Bond length $l$; (b) Bond angle $\theta_{1}$; (c) Axial bond angle $\theta_{2}$. }
\end{figure*}

\subsection{Effect of Strain on the Magnetic Anisotropy Energy}
One can also observe how MAE changes with biaxial strain in Fig.~5. In Figs. 5(a)-5(c), the energy was calculated as a function of the angle of magnetization with respect to the basal plane ($\theta_M$), being 0º in-plane and 90º off-plane. This is illustrated in the inset of Fig.~5(f), where $E_\parallel $ and $E_\perp $ represent the energies calculated when all the spins are parallel and perpendicular to the atomic plane, respectively. These calculations were repeated for different values of strain. If we neglect the higher order terms, the dependency of the energy per chromium atom with respect to $\theta_M$ is given by: \cite{Bogdanov1998}
\begin{equation}
E(\theta_{M})=E_{0}+\lambda_{1}sin^{2}(\theta_{M})+\lambda_{2}sin^{4}(\theta_{M}).
\end{equation}
Here, $E_{0}$ is a constant energy shift, $\lambda_{1}$ and $\lambda_{2}$ are respectively the quadratic and quartic contributions to the energy. No substantial difference in energy with respect to the different in-plane directions are observed from DFT calculations, therefore the azimuthal contribution to $E(\theta_{M})$ is neglected. From Figs.~5(a)-5(c) we note that $E(\theta_{M})$ provides a good fit for the energies. In addition, it can be seen from Figs.~5(a)-5(c) that the easy axis remains off-plane for CrI$_{3}$ and CrBr$_{3}$ even when subject to strain, whereas for CrCl$_{3}$ there exists a phase transition to an in-plane easy axis upon compression. This can be further verified by looking at Figs.~5(d)-5(e), in which we take the difference between in-plane ($E_\parallel$) and off-plane ($E_\perp $) energies and plot them with respect to strain. Here, negative values seen in Fig.~5(c) for CrCl$_{3}$ represent an in-plane preference for magnetization.

Although bulk CrCl$_{3}$ is found to possess an in-plane easy axis \cite{McGuire20172}, it should be noted from Fig.~5(c) that the MAE of unstrained monolayer CrCl$_{3}$ is positive, indicating that the spins of Cr atoms align perpendicular to the basal plane, despite that the MAE is much smaller than that of CrI$_3$. Similar result was also reported by Zhang \emph{et al. }\cite{Zhang2015}. This result suggests a possible transition from an in-plane to an off-plane easy axis for CrCl$_{3}$ upon exfoliation. This is not surprising, considering that the MAE of CrCl$_3$ is much smaller than that of CrI$_3$. Furthermore, unlike CrBr$_{3}$ and CrCl$_{3}$ compounds, CrI$_{3}$ crystal exhibits much stronger anisotropy when compressed and becomes weaker when stretched. More specifically, a -5\% compressive strain will increase 47\% of MAE in this system.

Fitting parameters for $\lambda_{1}$ and $\lambda_{2}$ can be found in Figs.~5(g)-5(i). These plots show the strain dependence of these constants for the three monolayer systems, with the $\lambda_{2}$ graph being displayed in the inset. We note that, for all systems the quadratic contribution dominates the MAE, as the change of $\lambda_{1}$ resemble the MAE curves for each compound. However, as tensile strain increases, the quartic contribution gets comparable to the quadratic in magnitude for CrBr$_{3}$ and CrI$_{3}$.

Finally, we would like to point out that calculations based on LDA yield similar effects on strain on the MAE, although the value is slightly different (see Fig.~S3(a) in the Supporting Materials). In addition, the MAE is strongly dependent on the on-site parameter of Hubbard U for Cr. Our PBE+U calculations show that the MAE increases dramatically when the Hubbard U parameter is increased. The enhancement in the MAE with respect to a 5\% compressive strain ranges from 20.7\% - 58.1\% depending on the value of Hubbard U parameter. All the data has been included in the Supplemental Materials \cite{sup2018}.

\begin{figure*}[tbp]
\centering
\includegraphics[width=1.00\textwidth]{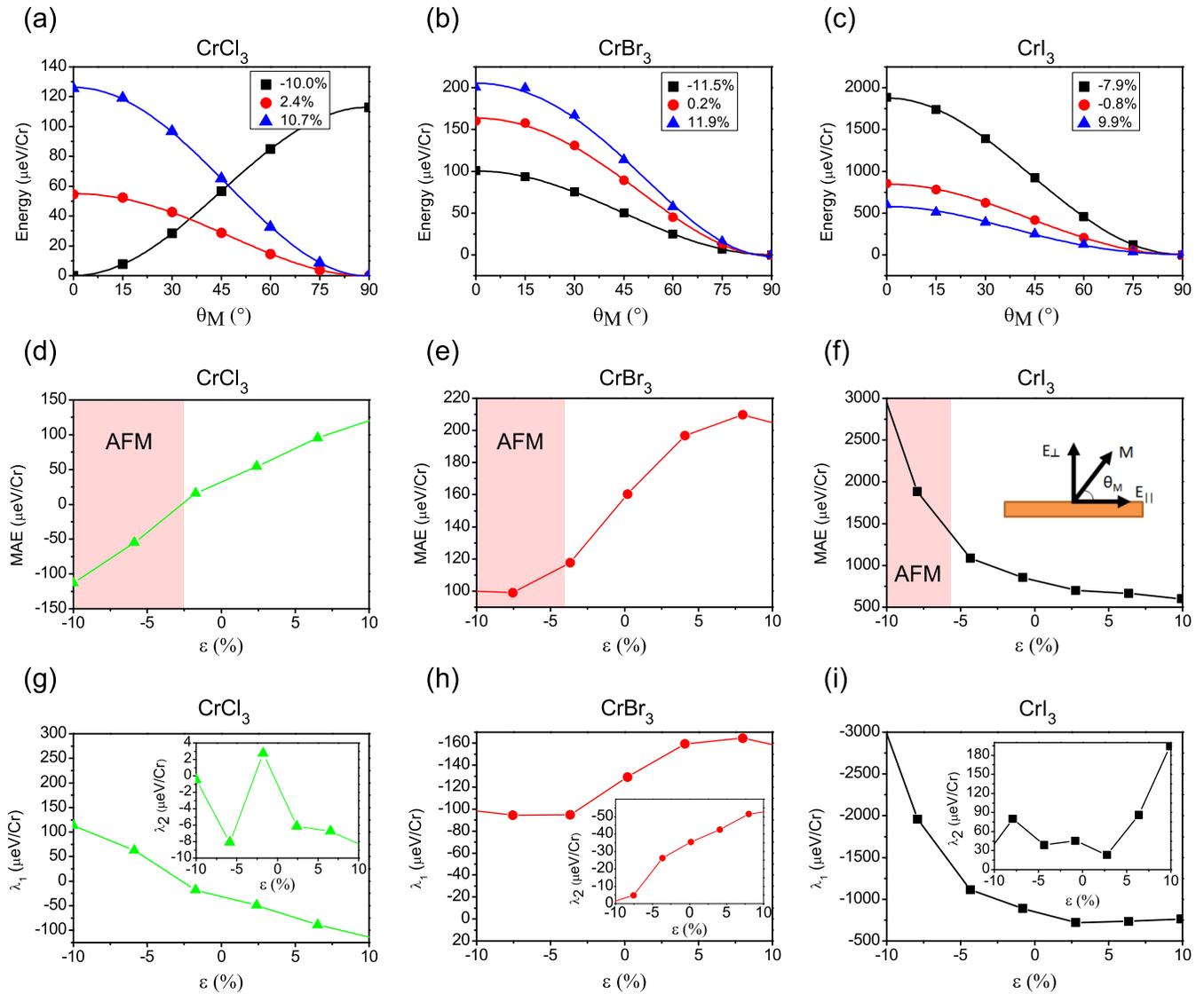}
\caption{\label{fig:band} The effect of strain on the magnetic anisotroy energy. Change of energy with respect to the magnetization angle $\theta_{M}$ for (a) CrCl$_3$, (b) CrBr$_3$ and (c) CrI$_3$. The line is the fitted result. (d)-(f) Change of MAE with respect to strain in (d) CrCl$_3$, (e) CrBr$_3$ and (f) CrI$_3$. The AFM phase region is highlighted in red. The change of the fitted parameters $\lambda_1$ and $\lambda_2$ are shown in (g)-(i).}
\end{figure*}

\section{Conclusions}\label{conclusion}
In summary, we applied density functional theory to investigate the magnetic and electronic properties of a recently found 2D magnet, monolayer CrI$_{3}$ and other compounds from the same family including CrBr$_{3}$ and CrCl$_{3}$. All three monolayer systems are found to be ferromagnetic in the ground state in accordance to previous work. In addition, CrI$_{3}$ exhibits strong magnetic anisotropy (804 $\mu$eV/Cr) with an easy axis perpendicular to the basal plane. The MAE decreases dramatically as the atomic number of the halide decreases, with CrBr$_{3}$ and CrCl$_{3}$ having 160 and 25 $\mu$eV/Cr respectively. We also estimated the Curie temperature from mean field approximation and the calculated energy differences between two different magnetic orders, namely FM and AFM. Our results are in good agreement with experimental data, and suggest a relatively large interlayer coupling in the CrI$_{3}$ system. Interestingly, our mean-field estimation predicts that the Curie temperature for CrCl$_{3}$ can increase from 29.7 K to 39 K with 2.4 \% of tensile strain.

Furthermore, we have studied the effect of biaxial strain and have determined the strain dependence of the atomic structure, electronic and magnetic properties of the chromium trihalides. Strong SOC in the CrI$_{3}$ results in a more evident strain dependence of the electronic and magnetic properties as compared with CrBr$_3$ and CrCl$_3$.  For example, the MAE in CrI$_{3}$ increases when compressed and decreases when stretched. A 5\% compressive strain will increase MAE by 47\%  in this system.

\section*{Acknowledgements}
This work used the Extreme Science and Engineering Discovery Environment (XSEDE) Comet at the SDSC through allocation TG-DMR160101 and TG-DMR16088. We acknowledge support from the NSF grant DMR 1709781 and support from the Fisher General Endowment and SET grants from the Jess and Mildred Fisher College of Science and Mathematics at the Towson University.

\pagebreak
\appendix{
\textbf{\large Supporting Documents: Strain-tunable magnetic anisotropy in monolayer CrCl$_3$, CrBr$_3$, and CrI$_3$}
\setcounter{equation}{0}
\setcounter{figure}{0}
\setcounter{table}{0}
\setcounter{page}{1}
\makeatletter
\renewcommand{\theequation}{S\arabic{equation}}
\renewcommand{\thetable}{S\arabic{table}}
\renewcommand{\thefigure}{S\arabic{figure}}
\renewcommand{\bibnumfmt}[1]{[S#1]}
\renewcommand{\citenumfont}[1]{S#1}

\subsection{Electronic structures from collinear calculations}

In this session we focus on results of PBE calculations without SOC. Shown in Figs. S1(a)-(c) are the electronic band structures for the CrCl$_3$, CrBr$_3$ and CrI$_3$, respectively. In this plot, the red lines represent the minority spin states and the blue lines represent the majority spin states. Clearly, the band gaps range from 1.12 eV to 1.50 eV, increasing as we move from iodine to chlorine, and the slight change in the band topologies is sufficient to change the band gap type from indirect to direct in the CrCl$_3$. Moreover, we find from Fig.~S1 that the valence band and conduction band near band edges are completely spin-polarized.


\begin{figure*}[htbp]
\centering
\includegraphics[width=1.00\textwidth]{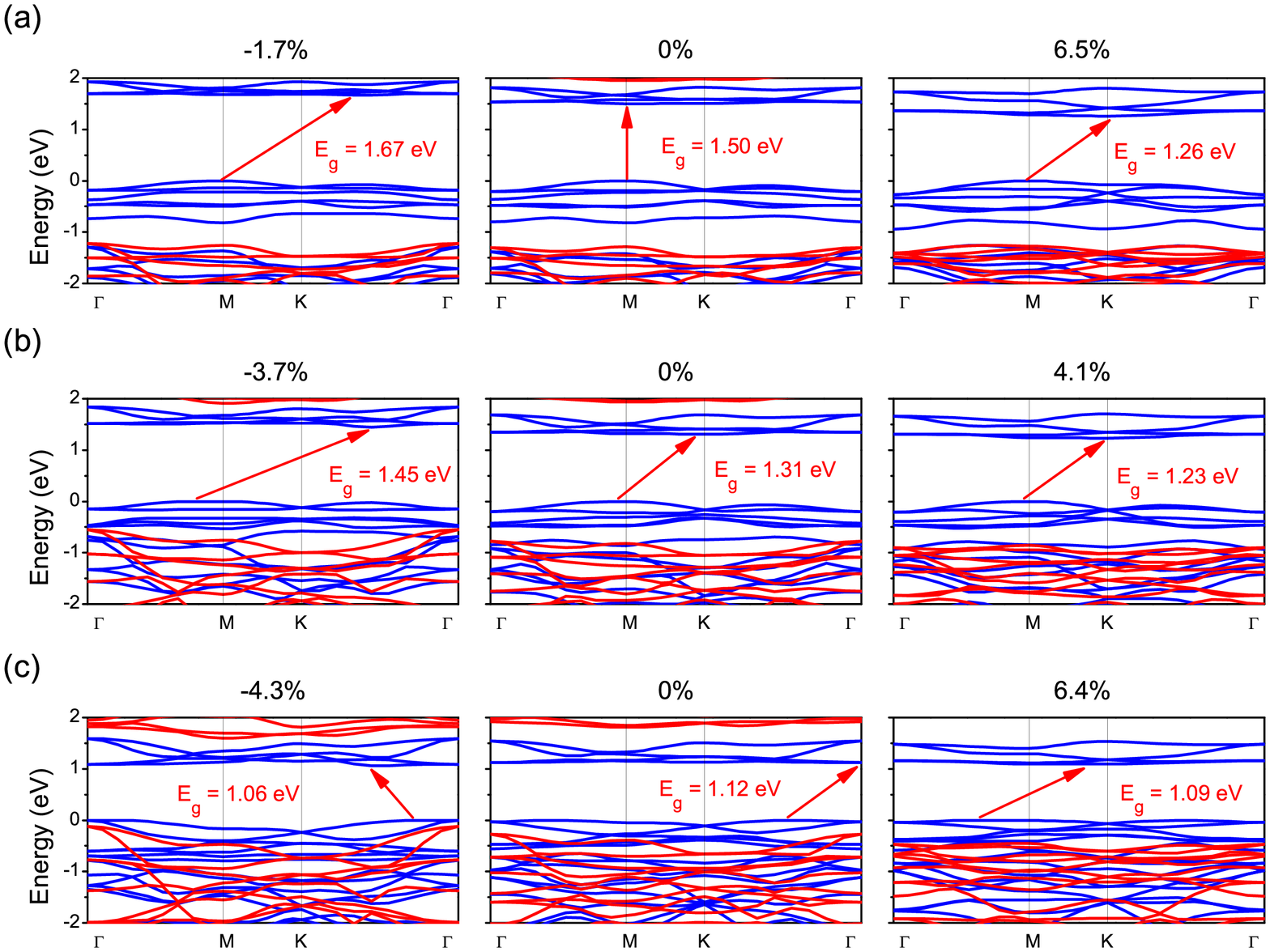}
\caption{\label{fig:band} Collinear spin-polarized electronic band dispersions for monolayer chromium trihalides. The effect of typical compressive and tensile strain are also shown in (a) CrCl$_{3}$, (b) CrBr$_{3}$, and (c) CrI$_{3}$. Red lines represent minority spin states, and blue lines represent majority spin states. The energy band gaps between conduction band minimum (CBM) and valence band maximum (VBM) have been indicated using red arrows in each case. The VBM has been shifted to zero.}
\end{figure*}

\subsection{Self consistent field approximation for estimating Curie temperature}
In this section we provide a short derivation of the Equation (4) used to estimate the Curie temperature for the chromium trihalides based on mean-field theory \cite{Hynninen2006,Fuh2016,Sato2003}. First, consider spins ($\sigma$) located in an hexagonal lattice, and let $\sigma=\pm1$. Then:

\begin{equation}\label{eqS1}
E=-J\sigma \sum_{neighbors} \sigma_{i},
\end{equation}
where $E$ is the energy per particle related to the spin of each atom, and $J$ is the spin coupling constant. According to the mean field approximation, we assume each site is coupled to an average spin that is due to the exchange coupling with its nearest neighbors. Then the energy can be written as:

\begin{equation}\label{eqS2}
E=-\frac{3}{2}J\sigma\bar{\sigma}.
\end{equation}
Here, the 3/2 factor comes from the fact that each atom in an hexagonal lattice has three nearest neighbors, and each unit cell contains two atoms. Next, we calculate the partition function for one particle:

\begin{equation}\label{eqS3}
Z=e^{3\beta J\bar{\sigma}/2}+e^{-3\beta J\bar{\sigma}/2}=2cosh(\frac{3\beta J\bar{\sigma }}{2}),
\end{equation}
where $\beta = 1/K_{B}T$. Then, the average Energy per particle can be determined by the following:

\begin{equation}\label{eqS4}
<E>=-\frac{1}{Z}\frac{\partial Z}{\partial \beta}=-\frac{3J\bar{\sigma }}{2}tanh(\frac{3\beta J\bar{\sigma }}{2}).
\end{equation}
From Eq.~\ref{eqS2}, we can calculate the average spin per particle:

\begin{equation}\label{eqS5}
<\sigma>=-\frac{2}{3J\bar{\sigma }}<E>=tanh(\frac{3\beta J\bar{\sigma }}{2}).
\end{equation}
Next, we make a self consistent approximation, in which we assume $<\sigma>=\bar{\sigma}$:

\begin{equation}\label{eqS6}
\bar{\sigma }=tanh(\frac{3\beta J\bar{\sigma }}{2}),
\end{equation}
\begin{equation}\label{eqS7}
\frac{2}{3\beta J}X=tanh(X),
\end{equation}
where $X\equiv \frac{3\beta J\bar{\sigma }}{2}$. This transcendental equation can be solved graphically, as shown in Fig.~S2, where we plotted the left and right hand side of Eq.~S7 separately. We can expect non-trivial solutions only when the slope of the line in the left hand side is sufficiently small, so that the two curves can cross at non-zero points. Specifically, the slope of the line at $X$=0 must be smaller than the slope of $tanh(X)$ at the same point. Hence, magnetic order will exist when the slope of the line in the left hand side of Eq.~\ref{eqS7} is smaller than 1. Fig.~S2 displays the phase transition limit, in which:

\begin{equation}
\frac{2}{3\beta J}=1,
\end{equation}
\begin{equation}
\frac{2K_{B}T_{c}}{3J}=1,
\end{equation}

\begin{equation}
T_{c}=\frac{3J}{2K_{B}}.
\end{equation}

\begin{figure*}[htbp]
\centering
\includegraphics[width=0.4\textwidth]{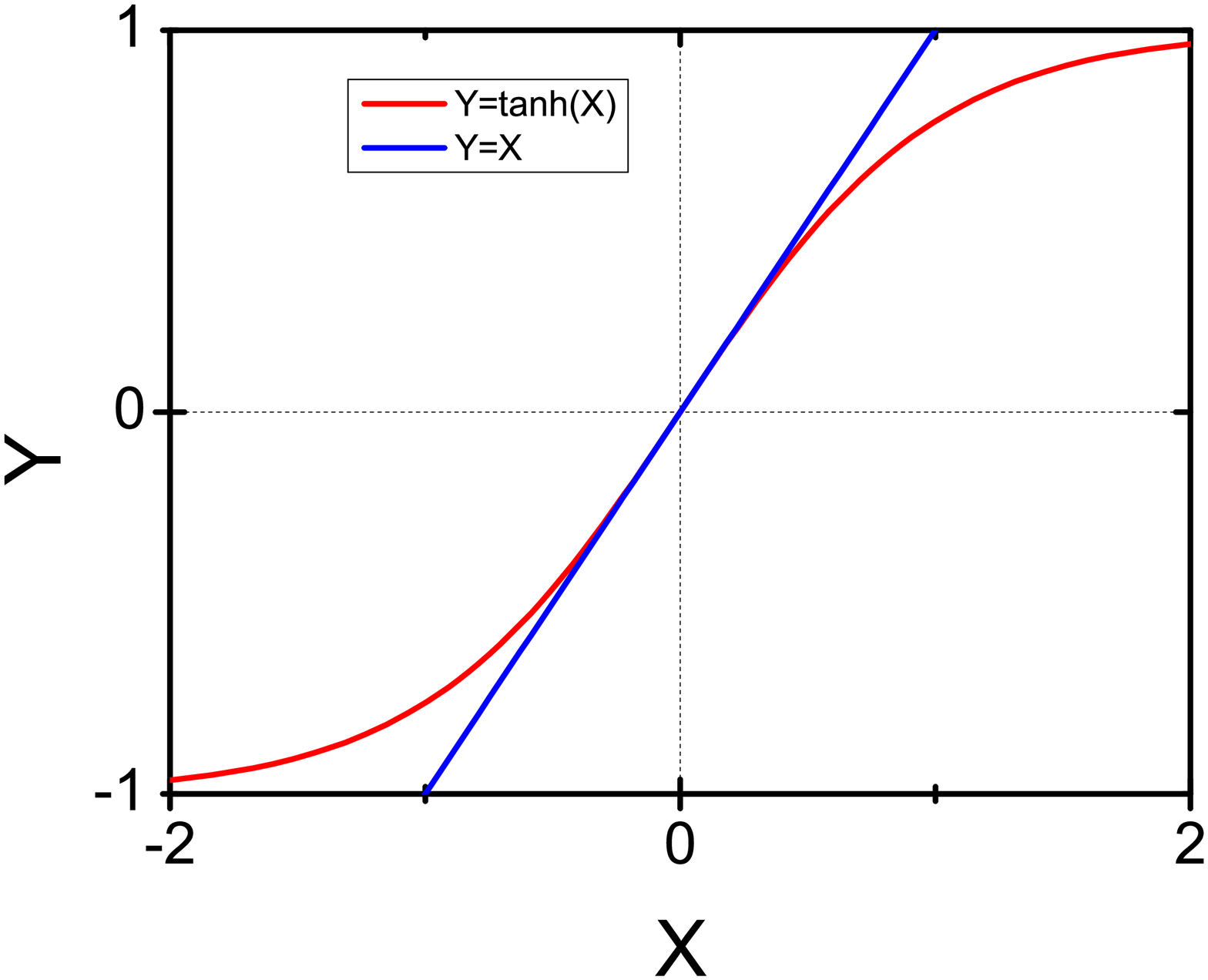}
\caption{\label{fig:trans} Graphical solution to the transcendental equation X=tanh(X)}
\end{figure*}

QED.

\subsection{Effect of exchange correlation and on-site Coulomb interactions on the magnetic anisotropy
energy}

We have also carried out calculations of CrI$_3$ using LDA to check the effects of exchange correlation functionals on the MAE. The results of LDA and PBE have been shown in Fig.~\ref{fig:U}(a). Overall, the two functionals yield nearly the same trend. Fig.~\ref{fig:U}(b) shows the results of CrI$_{3}$ calculated using PBE+U with U=1, 2.5 and 5 eV, respectively. It can be seen that the MAE depends on the parameter of U, which denotes the on-site Coulomb interactions for Cr. However, the trend of strain effect is similar: a compressive strain will increase the MAE. Furthermore, the increase in the MAE with respect to a 5\% compressive strain ranges from 20.7\% - 58.1\% depending on which exchange correlation functional is selected (LDA or PBE) or the value of Hubbard U parameter.

\begin{figure*}[htbp]
\centering
\includegraphics[width=1.0\textwidth]{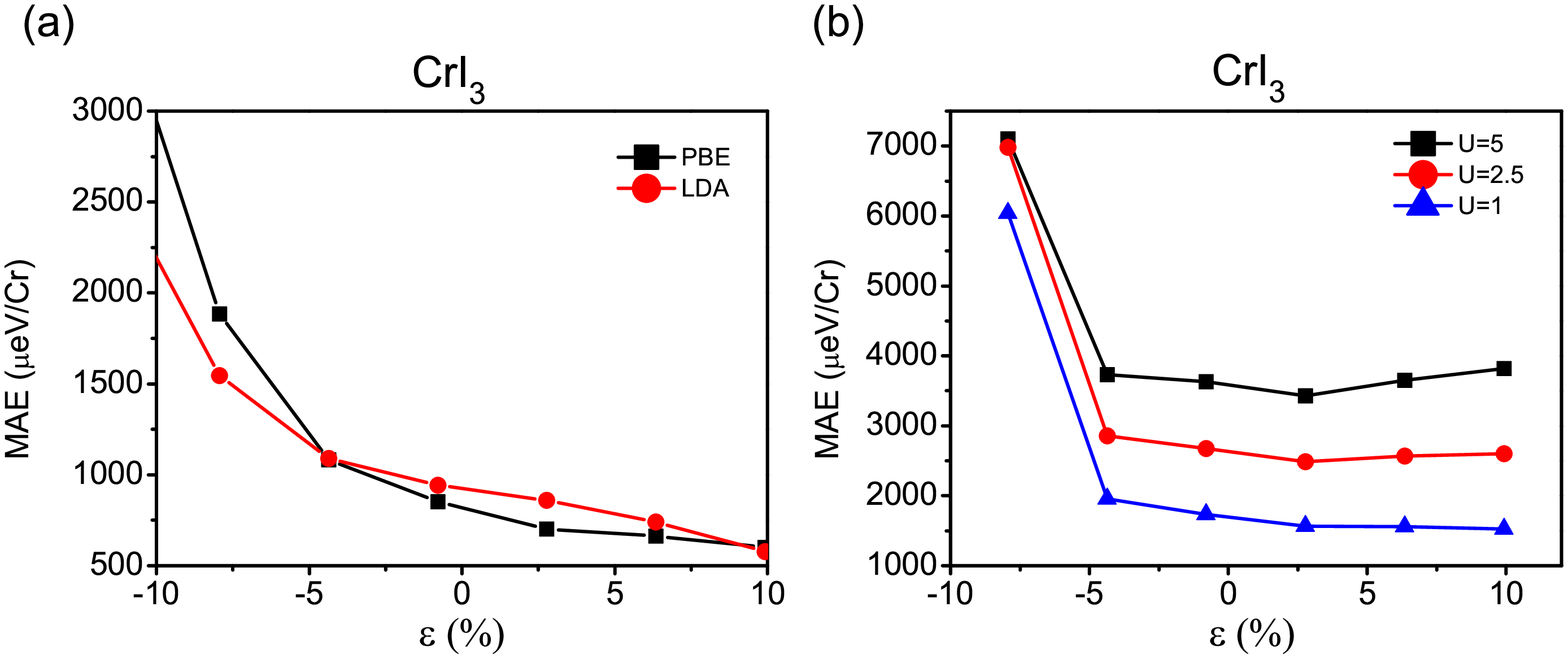}
\caption{\label{fig:U} Effects of strain on the magnetic anisotropy energy in CrI$_3$ calculated using (a) LDA and PBE, and (b)  PBE+U (with different U parameter).}
\end{figure*}

}

\end{document}